# Sparsity of weighted networks: measures and applications

Swati Goswami[ab*], Asit K. Das[a], Subhas C. Nandy[b]

[a]Department of Computer Science and Technology, Indian Institute of Engineering Science and Technology, Shibpur, Howrah – 711103, India

[b]Advanced Computing and Microelectronics Unit, Indian Statistical Institute, Kolkata, India, Pin 700108

**Abstract**

A majority of real-life networks are weighted and sparse. The present article aims at characterization of weighted networks based on sparsity, as an indicator of inherent diversity of different network parameters. The measure called *sparsity index* defined on ordered degree sequence of simple networks is extended and new properties of this index are derived. The range of possible values of sparsity index of any connected network, with edge-count in specific intervals, is worked out analytically in terms of node-count and a pattern is uncovered in corresponding degree sequences. Given the edge-weight frequency distribution of a network, an expression of the sparsity index of edge-weights is formulated. Its properties are analyzed under different distributions of edge-weights. For example, the upper and lower bounds of sparsity index of edge-weights of a network, with all distinct edge-weights, is determined in terms of its node-count and edge-density. The article highlights that this summary index with low computational cost, computed on different network parameters, is useful to reveal different structural and organizational aspects of networks for performing analysis. An application of this index is demonstrated through devising a new overlapping community detection method. The results validated on artificial and real-world networks show its efficacy.

**Keywords**

Sparsity Index, Sparsity Measure, Gini Index, Edge Weight Distribution, Overlapping Community Detection, Community Detection in Weighted Networks

## 1. Introduction

Network graphs are often used to model relationships among a group of interacting entities in the physical world. The entities themselves are considered as nodes, an interaction between two entities is represented by an edge and the intensities of interactions are represented by edge-weights. There are many real life examples of such networks, and they arise from different walks of human existence. The social [19, 48], biological [16, 45], cognitive [4, 41] relationships in human populations, the behavioral relationships among living creatures [36], the economic and trade relationships among nations or other territories [25], are examples of areas in which networks have been used extensively for research. A network is unweighted, when the model focuses only on the existence of a relationship between a pair of entities and ignores its intensity. Similarly, a network is undirected, when the direction of the relationship between two entities is of no significance for the model. The networks that are of specific interest to us, for the present article, are the weighted networks. In the network science literature of late, more emphasis is being given on the study of weighted networks. Weighted networks tend to mimic different real-world phenomena more closely than their unweighted counterparts, by attaching importance to the strength or magnitude of interactions at the modeling stage. In analyzing many networks that are part and parcel of modern human civilization, for example, transportation networks [5, 47], communication networks [27, 31], web networks [6, 42] and so on, weighted networks find more appropriate use. Also, it is a common observation that large-scale real life networks are highly sparse. Therefore, studying the network property of sparsity, in the case of weighted networks, is worthwhile.

For a network graph, sparsity reveals the extent of the graph's deviation from the corresponding fully connected graph. Sparsity lies at the opposite end of the spectrum of density of the graph. However, the term sparsity carries a different meaning when applied in the context of distribution of a certain quantity among a group of entities. In that context, sparsity indicates the relative diversity in distribution of a certain resource among related entities. For a network, we may be interested to measure how *sparsely* distributed the total degree (degree of a node is the number

---

[*] Corresponding Author. E-mail address: swati.goswami2000@gmail.com; Tel: +91 9836360340; Fax: +91 33 25773035

of edges incident on that node) of the graph is among its constituent nodes. This question assumes tremendous importance from various network analysis viewpoints. For example, in studying the propagation of failure through a system of inter-dependent networks [23], it has been shown that systems having degree distributions with higher variability tend to be more robust to random failures. From a community detection viewpoint, a network scoring low on sparsity of its node-degrees is unlikely to yield good communities. For dynamical processes like disease spreading, rumor spreading, opinion formation and evolution of social behaviors in networks, network topology has a fundamental role to play and sparsity of degree distribution, or *degree heterogeneity,* is of particular importance [35]. *Sparsity index* [18] is a summary measure based on the number of nodes, the degrees of the nodes and a constant factor at least as large as the total degrees of all nodes of the network. It indicates the extent of heterogeneity in distribution of degrees among the constituent nodes of the network. With a suitable choice of the constant factor as mentioned in its definition, the measure provides an indication of the graph's deviation from the corresponding complete graph. Notably, this measure applies equally on weighted and unweighted networks, as it concerns itself primarily with degrees of nodes of the network. In this article we have delved deeper into properties of this index and have shown analytically that the sparsity index attains the highest value for a star graph and the lowest value for a Hamiltonian path among connected graphs with the same number of nodes and edges. This is a reflection of the way sparsity index addresses the structural aspect of heterogeneity associated with complex networks. The limiting value of sparsity index is found to be ½ for an infinite star network. Further, a pattern has been uncovered in degree sequences of networks to produce maximum sparsity with specific edge-counts.

On a similar note, one may also be interested to find out a measure of sparsity of distribution of edge weights of a network. Edge-weights denote the intensities or the frequencies of connections; to what extent they vary in a network is a network property which can be exploited to analyze other properties and explain behavior of the network. For instance, in formulating a measure of complexity of brain networks, namely Shannon graph complexity [17], Gomez-Pilar et al. make use of the weight distribution in hypothesizing that for a fixed topology, the weight distribution of a complex network is directly related to its reliability and information propagation speed. In bioinformatics domain, Apeltsin et al. [1] have shown that pre-filtering a protein similarity network, based on a certain threshold on edge-weights, leads to an improvement in performance of clustering algorithms meant to group functionally similar proteins. The study further brings out a relationship between one such optimal threshold and the shape of the edge-weight distribution of the input protein similarity network. In the telecom domain, Dasgupta et al. [9] examine how social ties impact the propensity to churn in mobile telecom networks. Based on diffusion model, their work brings out the role of strong and weak ties (tie strength is expressed as edge-weight) in spreading the influence through the network. Interestingly however, all these different scenarios display a power-law distribution for the edge weights, with exponents lying somewhere between 1 and 3. Therefore, the nature of the distribution of edge weights is an important consideration for analysis of weighted networks and this may lead to exploration of other characteristics of the network.

It may be noted that the *degree* equivalent in a weighted network is *strength* [28]. The strength of a node is defined as the total weight of all edges incident on that node. Conceptually, strength represents an amalgamation of a node's degree and associated edge-weight. For weighted networks, it is legitimate to ask for a measure of sparsity of the strength distribution of its nodes. For example, a community detection process in general, which takes into account edge-weights in some way to form the communities, may benefit from using this measure. A low value of this measure may straightaway indicate that well-defined communities may not be present at all within the network. Thus, a sparsity measure can be constructed on various network parameters and be computed to throw light on the extent of diversity present in that specific aspect of the network. The computational cost is insignificant (to the order of $nlogn$) to compute these summary measures even for large networks, but the knowledge of the organization and structure of the network that is obtained through these measures during initial examination, may be significant.

From information theoretic viewpoint, a measure which is often associated with degree distribution of complex networks is Shannon *entropy*. The definition of entropy based on a random walk model of random traversal through the network, quantifies the heterogeneity in the network's degree distribution, in a sense that nodes with low degrees lower the overall network entropy, while high degree hubs cause its increase [44]. An interpretation of entropy, according to Wiedermann et al., is a measure of regularity or order in the network with respect to its navigability that is measured in terms of random walk. Just as sparsity index is a normalized measure, so is entropy, whereas the two approaches towards measuring diversity are very different and one is not quite a substitute of the other.

To demonstrate an application of the sparsity indices discussed in his article, we have chosen the area of community detection. The proposed method identifies the central nodes to form communities based on the sequence of node-strength of the network and thereby it makes use of the edge-weight information of the network. The method follows the overall structure of the overlapping community detection method as proposed by Goswami et al. [18] for simple networks. While the method for simple networks uses sparsity index defined on ordered degree sequence, the proposed method uses sparsity index defined on ordered strength sequence of the network. At the candidate community formation stage the new method makes use of the sparsity index on edge-weights. Notably, the method does not require any input parameter in terms of number or size of the communities and is focused on uncovering natural groupings of nodes that are inherent in the network.

A common approach to study network properties, or, measures reflecting different network characteristics, is to uncover them first for simple networks and extend them, later, for weighted networks. For example, the centrality measure *betweenness* [11] was first introduced for simple networks and was later used by Park et al. [33] to characterize weighted networks. The authors analyze how this centrality measure relates to the degrees and the weights of a network. Another instance is the work by Saramaki et al. [38], which deals with several generalizations of another important network parameter *clustering coefficient*, for weighted networks. Further, Garas et al. [12] provide a generalization of a method *k-shell decomposition*, commonly used to identify the most influential nodes in terms of spreading in networks, for weighted networks. In this article, we follow similar approach, by extending a measure of sparsity originally proposed for simple networks to weighted networks.

The main contribution of this paper is three-fold: discovering new properties of sparsity index based on degree sequence of networks; formulating sparsity index of edge weights of networks; demonstrating an application of the index by devising a community detection algorithm. Towards this end, we have shown analytically that the sparsity index based on ordered degree sequence of a connected graph with $n$ ($\geq 3$) nodes and $(n-1)$ edges lies between the sparsity indices of a Hamiltonian path and a star network of the same number of nodes and edges. While doing so, the progression of the graphs to attain the lowest to the highest values of this index has been brought out by using the corresponding degree sequences. Also, a general version of the result has been included, in the sense that the condition on edge-counts has been relaxed and the upper and lower bounds of sparsity indices have been worked out in terms of sequential intervals of edge-counts, such that any connected graph can fit in one of the intervals. This has been done in section 2 of this document. Given the frequency distribution of edge weights of a connected network, a formulation of the sparsity index of edge weights has been obtained directly from the corresponding Lorenz curve [13] definition and the index itself has been studied under different edge-weight frequency distributions. This has been done in section 3. A node-strength based community detection algorithm has been proposed for weighted graphs. This method utilizes sparsity index on node-strengths, Gini index of node-degrees and sparsity index on edge-weights at different stages of the process and follows the overall structure of an existing algorithm meant for simple networks [18]. The method has been applied on a synthetic network and a couple of benchmark real life networks for validation. The community detection method and the related experimental results appear in section 4 of this document. In section 5, the article has been concluded after summarizing the implications of the results obtained, both theoretically and experimentally, with a discussion on directions in which future work may proceed.

## 2. Sparsity Measure of network graphs

By applying a sparsity measure on a network graph, weighted or unweighted, one expects to get an idea of i) the extent to which it deviates from a fully connected network ii) the inherent variability among the degrees of nodes of the network. Sparsity index has the potential to throw light on both these aspects, by appropriately choosing the constant factor at least as large as the total degrees of all nodes, as used in its expression (refer to expression (1) in section 2.2). As we focus our attention specifically on weighted networks, we observe that sparsity as a measure can apply on different aspects of a weighted network, namely, the degrees and strengths of its nodes, edge-weights and so on. In this section we recapitulate the definition of sparsity index as it applies generally on the degree sequence of a network and explore the nature of graphs which yields the highest and the lowest values of this index, derive the index for the edge-weights of a weighted network given its frequency distribution and study the behavior of the index under specific scenarios. In section 2.1 we develop the preliminaries and introduce terms and notations which are to be followed across this article, in section 2.2 we review sparsity index and prove theorems regarding the maximum and minimum values of the index and in section 2.3 we construct the sparsity index applicable to edge-weights of a network.

## 2.1. Terms and Notations

Let the interactions among a group of $n$ entities be represented by a graph $G = (V, E)$ with $|V| = n$. If an interaction between nodes $i$ and $j$ has taken place, $i, j = 1,2, \ldots, n, i \neq j$, then there is an edge between them in the graph. We assume that the network topology of the input graph is known and the degree of each node can be easily computed. Let the adjacency matrix of the graph be denoted by $A = [a_{ij}]_{n \times n}, i, j = 1,2, \ldots, n; i \neq j$. For an unweighted graph, $a_{ij} \in \{0,1\}$ represents the presence of an edge between the nodes $i$ and $j$ and $a_{ii} = 0$. Let $a_i = \sum_{j=1}^{n} a_{ij}, i = 1,2, \ldots, n$ denote the number of interactions made by the $i$-th node of the graph. Let $T$ denote the sum of all interactions made by all of these $n$ entities, such that $\sum_{i=1}^{n} a_i = T$. The sequence $\underline{a} = \{a_1, a_2, \ldots, a_n\}$ represents the degree sequence of the graph. Let the elements of the degree sequence $\underline{a}$ be arranged in an increasing order and let $d_i$ denote the $i$-th ordered statistic in the sequence of values $\{a_1, a_2, \ldots, a_n\}$, i.e., $\underline{d} = \{d_1, d_2, \ldots, d_n\}$ such that $d_1 \leq d_2 \leq \cdots \leq d_n$. The sequence $\underline{d}$ is regarded as the ordered degree sequence of the graph.

If the magnitude of interactions among the entities can be quantified and assigned values, we get a weighted graph and $a_{ij}$ can take real values. Also, if there is no interaction between two entities, let the magnitude of the relationship be treated as zero. In other words, we denote the weight of a non-existing edge of the graph by zero, and consequently do not distinguish between an edge with zero weight-value and a non-existing edge. In a weighted graph, the sum of weights of incident edges on a specific node is called its *strength*; the strength of the node $i$ is given by: $s_i = \sum_{j=1}^{n} a_{ij}, i = 1,2, \ldots, n; a_{ij} \in \mathbb{R}$ and let the vector $\underline{s} = [s_1, s_2, \ldots, s_n]$ with $s_1 \leq s_2 \leq \cdots \leq s_n$ denote the ordered strength vector of the graph.

Let us now consider the edge-weight frequency distribution of the graph. Without loss of generality, let us assume that the edge-weights of the graph are non-negative integers. Let there be $k$ distinct weight values associated with the edges of the graph and the ordered weight sequence be represented by $\underline{w} = [w_1, w_2, \ldots, w_k]$, such that $0 \leq w_1 < w_2 < \cdots < w_k$. Let the corresponding frequencies of occurrence of these $k$ weights among $|E|$ edges be denoted by $f_1, f_2, \ldots, f_k$. Let the frequency distribution of edge-weights be denoted as in Table 1.

Table 1: Frequency Distribution of edge-weights of a network graph

| Weights ($w_i$) | Frequencies ($f_i$) |
|---|---|
| $w_1 \geq 0$ | $f_1$ |
| $w_2 > w_1$ | $f_2$ |
| . | . |
| . | . |
| . | . |
| $w_k > w_{k-1}$ | $f_k$ |
| Total | $\sum_{i=1}^{k} f_i = \binom{n}{2}$ |

According to our setting, $|E| = \binom{n}{2} = \sum_{i=1}^{k} f_i$. For, if there is any gap between the potential number of edges and actual number of edges in the graph, the weight $w_1 = 0$ fills that gap with frequency $f_1$. This implies that only for complete graphs $w_1$ can be strictly positive.

**Lorenz curve**: Given a set of n ordered numbers $d_1, d_2, \ldots, d_n$, the empirical Lorenz curve generated by them is defined at the points $\frac{i}{n}, i = 0, 1, \ldots, n$ by $L(0) = 0$ and $L\left(\frac{i}{n}\right) = \frac{c_i}{c_n}$, where $c_i = \sum_{j=1}^{i} d_j$. The empirical Lorenz curve, $L(p)$, is defined for all $p$ in the interval $(0,1)$ by linear interpolation and represents the fraction of the total variable measured that the holders of the smallest $p$-th fraction possess [13].

### 2.2. Sparsity index of simple networks

At the root of computing sparsity index of a graph, based on its ordered degree sequence, is the Lorenz curve. Also, according to the definition of sparsity index, the sparsity of a graph is measured with respect to a fixed quantity, say $T_1$, where $T_1 \geq T = \sum_{i=1}^{n} d_i$. Given the ordered degree sequence $\underline{d} = [d_1, d_2, \ldots, d_n]$ of a simple graph $G$ and the fixed quantity $T_1$, sparsity index, say $SI(G)$, is expressed as in equation 1 [18].

$$SI(G) = 1 - \frac{2}{nT_1}\sum_{i=1}^{n}(n - i + \tfrac{1}{2})d_i \tag{1}$$

$SI(G)$ is normalized, i.e., takes values in the range [0,1]. The value is zero for a complete graph, or, for any regular graph. Sparsity index takes higher values as more and more degrees are piled up on fewer nodes. We have constructed theorem 1 later in this section by specifying the maximum and minimum possible values the sparsity index can take for some special connected graphs. For simple graphs the most plausible choice of $T_1$ is $n(n-1)$, i.e., when it is computed with respect the *potential* total degrees of all nodes of the graph [18]. However, the choice of $T_1$ is guided entirely by the problem under consideration and the basis on which one wants to measure sparsity of the graph. For $T_1 = T$, the measure becomes exactly equal to Gini index [15] of the corresponding degree sequence.

Figure 1.*i*) represents the Lorenz curves drawn with two different choices of $T_1$, (1) with $T_1 = T$ and (2) with $T_1 > T$. It may be noted that for the second case, the curve meets the vertical line through (1,0) and (1,1) at a point lower than (1,1), denoted by $O'$ in the figure. Gini index is defined based on curve (1) and is equal to the ratio of the area bounded by the Lorenz curve and the line of equality (i.e., the shaded region in the figure, say A) to the area of the unit triangle containing the curve. If we call the area below the Lorenz curve as B, then Gini index is equal to $\frac{A}{A+B} = 1 - 2B$, since the area of the unit triangle is $\frac{1}{2}$.

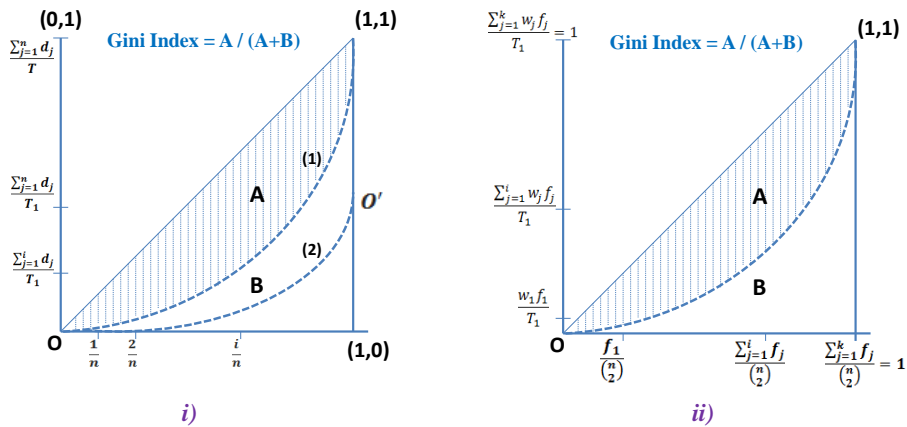

**Figure 1.** *i*) Lorenz curves for sparsity of degree distribution of nodes, drawn with two different choices of $T_1$, (1) with $T_1 = T$ and (2) with $T_1 > T$ *ii*) Lorenz curve to denote the sparsity of distribution of edge-weights of a weighted network.

With the definition of sparsity index in place, we explore the limiting values of this index for some special graphs and develop Theorem 1.

**Theorem 1:** With $n$ nodes ($n \geq 3$) and $n - 1$ edges, sparsity index of a connected network graph lies between $\frac{n-2}{n(n-1)}$ and $\frac{n-2}{2n}$.

To prove the theorem, we prove the following two lemmas:

Among connected graphs with $n$ nodes, $n \geq 3$ and $(n - 1)$ edges,

**Lemma 1:** a star graph with degree sequence $\{1, 1, \ldots \langle n - 1 \rangle \ times \ldots 1, (n - 1)\}$ attains the maximum sparsity index and

**Lemma 2:** a Hamiltonian path with degree sequence $\{1, 1, 2, 2, \ldots \langle n - 2 \rangle \ times \ldots 2, 2\}$ attains the minimum sparsity index.

As we proceed to prove the two lemmas, we draw reference to Figure 2, which demonstrates the degree sequences of graphs in increasing order of their sparsity indices. A clear pattern can be noticed in these degree sequences, and thereby in the corresponding graphs, which are essentially trees. Thus, through a specific example, the figure provides a footprint of how a Hamiltonian path evolves to a star graph, in terms of sparsity of their degrees.

Proof of **Lemma 1**:

Let the degree sequence $\{1, 1, \ldots \langle n - 1 \rangle \ times \ldots , 1, (n - 1)\}$ of a star graph with $n$ nodes be called as $S_1$. The sequence $S_1$ contains 1-s in the first $(n - 1)$ positions and $(n - 1)$ in the final position and is ordered. Let there be another degree sequence $S_2$, **ordered** and **distinct** from $S_1$, say, $S_2 = \{1, 1, \ldots \langle n - 2 \rangle \ times \ldots , 1, 2, (n - 2)\}$, in which there are 1-s in the first $(n - 2)$ positions, 2 in the $(n - 1)$-th position and $(n - 2)$ in the final position such that the total degree ($T$) of all nodes remains the same, i.e., $T = 2(n - 1)$, as in $S_1$. Note that $S_2$ in a sense gives the closest variation of the graph $S_1$, satisfying the condition of the theorem 1. The lemma is going to be proved for this specific case first and, afterwards, we move on to prove a more general case.

The theorem is true for the **specific case**, if and only if $SI(S_1) \geq SI(S_2)$     (2)

$SI(S_1) \geq SI(S_2) \Leftrightarrow 1 - \frac{2}{nT} [\sum_{i=1}^{n-1}(n - i + \frac{1}{2}) + \frac{n-1}{2}] \geq 1 - \frac{2}{nT} [\sum_{i=1}^{n-2}(n - i + \frac{1}{2}) + \frac{3}{2} 2 + \frac{n-2}{2}]$

$\Leftrightarrow -\frac{2}{nT} \sum_{i=1}^{n-1}(n - i + \frac{1}{2}) - \frac{n-1}{nT} \geq -\frac{2}{nT} \sum_{i=1}^{n-2}(n - i + \frac{1}{2}) - \frac{6}{nT} - \frac{n-2}{nT}$

$\Leftrightarrow \frac{2}{nT} \{\sum_{i=1}^{n-2}(n - i + \frac{1}{2}) - \sum_{i=1}^{n-1}(n - i + \frac{1}{2})\} + \frac{6}{nT} + \frac{n-2}{nT} - \frac{n-1}{nT} \geq 0 \Leftrightarrow \frac{3}{nT} - \frac{1}{nT} \geq 0$, which is true and thereby the expression (2) holds.

Let us consider the **general case**, as we define $S_2$ as $S_2 = \{1, 1, \ldots \langle n - k - 1 \rangle \ times \ldots , 1, a_{n-k}, a_{n-k+1}, \ldots, a_n\}$. $k$ is a positive integer, $1 \leq k < n - 1$, such that there is $k$ number of $a_i$-s with $a_i > 1$ and $1 < a_{n-k} \leq a_{n-k+1} \leq \cdots \leq a_n \leq (n - 1)$. Also, $\sum_{i=1}^{n} a_i = 2(n - 1)$, where $a_i = 1 \ \forall \ i = 1, 2, \ldots, n - k - 1$. So, for different values of the parameter, we get different degree sequences of $n$-terms. Thus, $S_2$ gives the general connected graph with $n$ nodes and $(n - 1)$ edges. In this general case,

$SI(S_1) \geq SI(S_2) \Leftrightarrow 1 - \frac{2}{nT} \{ \sum_{i=1}^{n-1}(n - i + \frac{1}{2}) + \frac{n-1}{2}\} \geq 1 - \frac{2}{nT} \sum_{i=1}^{n}(n - i + \frac{1}{2})a_i$

$\Leftrightarrow \sum_{i=1}^{n-k-1}(n - i + \frac{1}{2})(1 - a_i) + (n - (n - k) + \frac{1}{2})(1 - a_{n-k}) + (n - (n - k + 1) + \frac{1}{2})(1 - a_{n-k+1}) + \cdots + \frac{1}{2}(n - 1 - a_n) \leq 0$

$\Leftrightarrow (k + \frac{1}{2})(1 - a_{n-k}) + ((k - 1) + \frac{1}{2})(1 - a_{n-k+1}) + \cdots + (k - (k - 1) + \frac{1}{2})(1 - a_{n-1}) + \frac{1}{2}(n - 1 - a_n) \leq 0$

(the above step follows from the previous one, as we note that $a_i = 1 \ \forall \ i = 1, 2, \ldots, n - k - 1$)

$\Leftrightarrow (k + \frac{1}{2})(1 - a_{n-k}) + (k - \frac{1}{2})(1 - a_{n-k+1}) + \cdots + \frac{3}{2}(1 - a_{n-1}) + \frac{1}{2}(n - 1 - a_n) \leq 0$

$\Leftrightarrow (2k + 1)(1 - a_{n-k}) + (2k - 1)(1 - a_{n-k+1}) + \cdots + 3(1 - a_{n-1}) + (n - 1) - a_n \leq 0$ (3)

Now, $T = (n - k - 1) + a_{n-k} + a_{n-k+1} + \cdots + a_n = 2(n - 1) \Rightarrow a_n = 2n - 2 - n + k + 1 - a_{n-k} - a_{n-k+1} - \cdots - a_{n-1}$.

The expression (3) is equivalent of showing:

$(2k + 1)(1 - a_{n-k}) + (2k - 1)(1 - a_{n-k+1}) + \cdots + 3(1 - a_{n-1}) + (n - 1) - (n + k - 1 - a_{n-k} - a_{n-k+1} - \cdots - a_{n-1}) \leq 0$

$\Leftrightarrow (2k + 1)(1 - a_{n-k}) + (2k - 1)(1 - a_{n-k+1}) + \cdots + 3(1 - a_{n-1}) - (k - a_{n-k} - a_{n-k+1} - \cdots - a_{n-1}) \leq 0$

$\Leftrightarrow (2k + 1)(1 - a_{n-k}) + (2k - 1)(1 - a_{n-k+1}) + \cdots + 3(1 - a_{n-1}) - (1 - a_{n-k} + 1 - a_{n-k+1} + \cdots + 1 - a_{n-1})$

$\Leftrightarrow 2k(1 - a_{n-k}) + (2k - 2)(1 - a_{n-k+1}) + \cdots + 2(1 - a_{n-1}) \leq 0$.

We note that $a_i > 1 \; \forall \; i \geq n - k$, therefore each term in parenthesis of the above expression is negative and the expression is true. Therefore, the relationship (3) holds and $SI(S_1) \geq SI(S_2)$.

$SI(S_1)$ can be computed as: $SI(S_1) = 1 - \frac{2}{nT}\{\sum_{i=1}^{n-1}(n - i + \frac{1}{2}) + \frac{n-1}{2}\} = 1 - \frac{2}{nT}\{(n - 1)(n + \frac{1}{2}) - \frac{(n-1)n}{2} + \frac{n-1}{2}\}$

$= 1 - \frac{2}{n.2(n-1)}(n - 1)(n + \frac{1}{2} - \frac{n}{2} + \frac{1}{2}) = 1 - \frac{1}{n}(\frac{n+2}{2}) = \frac{n-2}{2n}$ (4)

Proof of **Lemma 2**:

To prove Lemma 2, we state and prove **Lemma 3** first and the proof continues afterwards.

**Lemma 3:** Let us consider the degree sequences $\{a\}$ and $\{b\}$, in general, satisfying the following:

i. $\{a\} = \{a_1, a_2, \ldots, a_n\}$, $n \geq 3$ be a sequence of positive integers such that $1 \leq a_1 \leq \cdots \leq a_n \leq (n - 1)$ and $\sum_{i=1}^{n} a_i = 2(n - 1) = T$, say.
ii. $\{b\} = \{b_1, b_2, \ldots, b_n\}$ be another sequence of positive integers such that $1 \leq b_1 \leq \cdots \leq b_n \leq (n - 1)$ and $\sum_{i=1}^{n} b_i = 2(n - 1) = T$.
iii. $b_i \leq a_i \; \forall \; i = 1,2, \ldots, \alpha$ and $b_i \geq a_i \; \forall \; i = \alpha + 1, \ldots, n$ where $1 \leq \alpha \leq (n - 1)$.

Then $SI(\{a\}) \leq SI(\{b\})$

Proof of **Lemma 3:**

Here, $SI(\{a\}) \leq SI(\{b\}) \Leftrightarrow 1 - \frac{2}{nT}\sum_{i=1}^{n}(n - i + \frac{1}{2})a_i \leq 1 - \frac{2}{nT}\sum_{i=1}^{n}(n - i + \frac{1}{2})b_i \Leftrightarrow \sum_{i=1}^{n} ib_i - \sum_{i=1}^{n} ia_i \geq 0$

$\Leftrightarrow \sum_{i=1}^{\alpha} i(b_i - a_i) + \sum_{i=\alpha+1}^{n} i(b_i - a_i) \geq 0$. (5)

Of the two terms in expression (5), the second term is non-negative and is associated with larger weights in terms of $i$. Also, both sequences are non-decreasing and differences $(b_i - a_i)$ eventually cancel each other so that their sum remains zero. In fact, the magnitude of differences $(b_i - a_i)$ to the left of $\alpha$ is equal to the same to the right of $\alpha$ (as otherwise the differences cannot cancel each other).

$\sum_{i=1}^{n}(b_i - a_i) = 0 \Rightarrow \sum_{i=1}^{\alpha}(b_i - a_i) = \sum_{i=\alpha+1}^{n} -(b_i - a_i) \Rightarrow \sum_{i=1}^{\alpha}|b_i - a_i| = \sum_{i=\alpha+1}^{n}|b_i - a_i| \Rightarrow \sum_{i=1}^{\alpha} i|b_i - a_i| \leq \sum_{i=\alpha+1}^{n} i|b_i - a_i|$. Therefore, the magnitude of the non-negative term in expression (5) is higher and the relationship holds. Thus, we have shown that $SI(\{a\}) \leq SI(\{b\})$ and have proved Lemma3. ∎

Now, let us observe the sequences carefully. A realization of the sequence $\{a\}$ can be of the form $\{1, 1, 2, 2, ... \langle n - 2\rangle \ times\ ..., 2\}$, let us call this as $S_1$. We also note that the sequence $\{a\}$ has to contain 1 as the first two elements in order to be a non-decreasing degree sequence of a connected graph with the total degree of $2(n - 1)$ (for, if not, it violates the total degree, or, the connectedness assumption). In order to obtain a realization of the sequence $\{b\}$, say $S_2$, we are to distribute $T = 2(n - 1)$ total degrees among $n$ nodes in such a way that the resulting degree sequence is a valid degree sequence of a connected graph and is non-decreasing. A combination of only 1-s and 2-s as in $S_1$ is not possible to construct $S_2$, distinct from $S_1$ (as otherwise it leads to a contradiction in terms of total degrees $T$). Therefore $S_2$ has to contain at least one element larger than 2. Let us construct $S_2$ in such a way that it contains exactly one element larger than 2, i.e., exactly one 3. Thus, the closest $S_2$ can get of $S_1$ is by having $S_2 = \{1, 1, 1, 2, 2, ... < n - 4 > times ..., 2, 3\}$. We note that $S_2$ satisfies all conditions of the sequence $\{b\}$, with $\alpha = 3$.

Also note that if we reverse the two sequences for $\{a\}$ and $\{b\}$, i.e., make $S_2 \equiv \{a\}$ and $S_1 \equiv \{b\}$, the condition of constructing $\{b\}$ is violated. In fact, the only possible choice of the sequence $\{a\}$ is $S_1$ and the result $SI(S_1) \leq SI(S_2)$ follows in whichever way we may choose to construct $S_2 \equiv \{b\}$. If there are a larger number of elements in $S_2$ with value 3 (or higher), the larger number of 2-s are to be changed to 1-s and the resulting sparsity is going to be higher. An illustration of possible construction of $S_2$ is provided in Figure 2, which shows degree sequences in increasing order of their sparsity indices.

Note that $S_1$ represents the degree sequence of a Hamiltonian path and it gives the least sparsity index among all connected graphs with the same number of nodes and edges.

$SI(S_1)$ can be computed as: $SI(S_1) = 1 - \frac{2}{nT}\{(n - \frac{1}{2}) + (n - \frac{3}{2}) + \sum_{i=3}^{n}(n - i + \frac{1}{2})2\} = 1 - \frac{2}{2n(n-1)}\{2n - 2 + \sum_{i=3}^{n}(n - i + \frac{1}{2})2\} = 1 - \frac{1}{n(n-1)}\{2n - 2 + 2(n - 2)(n + \frac{1}{2}) - 2\{\frac{n(n+1)}{2} - 3\}\} = 1 - \frac{2}{n} - \frac{1}{n(n-1)}\{(n - 2)(2n + 1) - n^2 - n + 6\} = 1 - \frac{2}{n} - \frac{1}{n(n-1)}(n - 2)^2 = \frac{n-2}{n(n-1)}.$ (6)

Combining relationships (4) and (6) from the two lemmas, it can be concluded that the degree sparsity index of a star graph attains the highest value and that of a Hamiltonian path attains the lowest value among all connected graphs $G$ with $|V| = n \geq 3$ and $|E| = n - 1$ and $\frac{n-2}{n(n-1)} \leq SI(G) \leq \frac{n-2}{2n}$. ∎

From this relationship, it follows that for very large $n$, the sparsity index of a star graph approaches ½ and that of a Hamiltonian path approaches 0.

| | Degree Sequences | | | | | | | Graphs |
|---|---|---|---|---|---|---|---|---|
| $S_1$ | 1 | 1 | 2 | 2 | ... <n-2> times ... | | 2 | 2 |
| $S_{2,1}$ | 1 | 1 | 1 | 2 | ... <n-4> times ... | | 2 | 3 |
| $S_{2,2}$ | 1 | 1 | 1 | 1 | 2 ... <n-6> times ... 2 | | 3 | 3 |
| $S_{2,3}$ | 1 | 1 | 1 | 1 | 2 ... <n-6> times ... 2 | | 2 | 4 |
| ⋮ | | | | | | | | |
| $S_{2,r-1}$ | 1 | 1 | 1 | 1 | 1 ... <n-2> times ... 1 | | 2 | (n-2) |
| $S_{2,r}$ | 1 | 1 | 1 | 1 | ... <n-1> times ... | | 1 | (n-1) |

**Figure 2.** Degree sequences of connected graphs with n=7 nodes and (n-1) edges in increasing order of their sparsity indices. A pattern can be noticed in the degree sequences and the corresponding graphs.

**Discussion:** Theorem 1 can be extended by relaxing $|E|$. As a next step to this relaxation process, we can take $|E|$ to satisfy $(n-1) \leq |E| \leq (n-1) + (n-2)$, $n \geq 3$. With $|E| = (n-1) + (n-2)$, i.e., $T = 2(n-1+n-2) = 4n-6$, the highest sparsity is attained by a graph with degree sequence, say $S_1 = \{2, 2, ... \langle n-2 \rangle \, times \, ..., 2, (n-1), (n-1)\}$, among connected graphs with the same number of nodes and edges. This follows directly from Lemma 3. [Proof: Suppose $S_1$ is not the degree sequence with highest sparsity. Therefore, there must be a sequence $S_2$ whose sparsity can be higher than $S_1$, yet it is an ascending sequence preserving the total degree and the connectedness constraints under consideration. A possible candidate $S_2$ with $SI(S_2) > SI(S_1)$ is $S_2 = \{1, 2, 2, ... \langle n-4 \rangle \, times \, ..., 2, 3, (n-1), (n-1)\}$. But this fails to be a degree sequence of a feasible graph [20]. Thus, changing any one of the leading $(n-2)$-many 2-s of $S_1$ to a lower value, or 1, leads to invalid graph. But, changing any of these 2-s to a value greater than 2 leads to a decrease in sparsity index, which is not our intention. Changing the last two elements of $S_1$ to a value lower than $(n-1)$, also leads to lower sparsity index. Therefore, it is not feasible to construct $S_2$ as we have set out to.].

$SI(S_1) = 1 - \frac{2}{n(4n-6)} [(n-\frac{1}{2})2 + (n-\frac{3}{2})2 + \cdots + \frac{3(n-1)}{2} + \frac{n-1}{2}] = \frac{(n-2)(n-3)}{n(2n-3)}$, after simplification.

We can formalize this result in terms of the below corollary.

**Corollary 1:** For a connected graph $G$ with $n \geq 3$ nodes and $|E|$ edges, where $(n-1) \leq |E| \leq (n-1) + (n-2)$, the sparsity index $SI(G)$ satisfies the relationship $\frac{(n-2)(n-3)}{n(2n-3)} \leq SI(G) \leq \frac{n-2}{2n}$.

Extending further, the next interval in terms of edge-count that can be considered is: $(n-1) + (n-2) \leq |E| \leq (n-1) + (n-2) + (n-3)$. A similar bound in terms of sparsity index can be worked out for this case as well. With $|E| = (n-1) + (n-2) + (n-3)$, i.e., $T = 6(n-2)$, the highest sparsity is attained by a graph with degree sequence $\{3, 3, ... \langle n-3 \rangle \, times \, ..., 3, 3, (n-1), (n-1), (n-1)\}$. The sparsity index of the network with this degree sequence is calculated as $\frac{(n-3)(n-4)}{2n(n-2)}$. We choose to update corollary 1 to include this result, instead of creating a new corollary.

**Corollary 1 (updated):** For a connected graph $G$ with $n \geq 3$ nodes and $|E|$ edges,

(i) if $(n-1) \leq |E| \leq (n-1) + (n-2)$, the sparsity index $SI(G)$ satisfies the relationship $\frac{(n-2)(n-3)}{n(2n-3)} \leq SI(G) \leq \frac{n-2}{2n}$ and

(ii) if $(n-1) + (n-2) \leq |E| \leq (n-1) + (n-2) + (n-3)$, the sparsity index $SI(G)$ satisfies the relationship $\frac{(n-3)(n-4)}{2n(n-2)} \leq SI(G) \leq \frac{(n-2)(n-3)}{n(2n-3)}$.

Continuing this way, for a graph $G$ with $n \geq 3$ nodes and edge-count $|E| = (n-1) + (n-2) + \cdots + (n-k); 1 \leq k \leq (n-1)$, and $T = k(2n - k - 1)$, the highest sparsity is attained by the degree sequence $\{k, k, \ldots \langle n - k \rangle \text{ times} \ldots, k, (n-1), \ldots \langle k \rangle \text{ times}, \ldots, (n-1)\}$ and the corresponding sparsity index can be computed as $\frac{(n-k)(n-k-1)}{n(2n-k-1)}$.

It is interesting to note that, with increasing edge-count sequentially as above, we end up having a complete graph. The expression $(n-1) + (n-2) + \cdots + (n - (n-2)) + (n - (n-1))$ is nothing but $\binom{n}{2}$, which is the edge-count of a complete graph. We recollect that the sparsity index of a complete graph is zero. Therefore, by taking sequential intervals of edge-counts, we gradually move to denser graphs and cover the entire spectrum as we reach the densest graph, which is a complete graph. As the result stated in (updated) Corollary 1 is extended to include the generalized version in terms of $k$, we write it in the form of Theorem 2 as follows. Additionally, for a diagrammatic representation of the result, we attach Figure 3.

**Theorem 2:** For a connected graph $G$ with $n \geq 3$ nodes and $|E|$ edges, such that
$(n-1) + (n-2) + \cdots + (n-k) \leq |E| \leq (n-1) + (n-2) + \cdots + (n-k) + (n-k-1); k = 1, 2, \ldots, (n-2)$,
the sparsity index $SI(G)$ satisfies the relationship $\frac{(n-k-1)(n-k-2)}{n(2n-k-2)} \leq SI(G) \leq \frac{(n-k)(n-k-1)}{n(2n-k-1)}$.

Let us observe Figure 3. For any connected graph $G$ of size $n \geq 3$, the edge-count $|E|$ either matches an edge-count value as shown in the "edge count" column in figure 3, or, falls in one of the intervals in between two successive edge-count values (as stated in Theorem 2). The sparsity index of graph $G$, $SI(G)$, finds its position in the corresponding interval of two successive highest sparsity values as shown in the fourth column of the same diagram. If one carefully examines the degree sequences as specified in the third column, a pattern can be observed, which is more vividly reflected in the corresponding example graphs in the last column. It must be noted that the graphs shown in the last column are only a set of example graphs for illustration purpose, assuming a fixed number of nodes.

| | Edge Count | Degree Seq. to yield highest sparsity (no. of times is mentioned in angular brackets) | Value of highest sparsity | Graph (example with n=7) |
|---|---|---|---|---|
| 1. | (n-1) | {1, 1, 1, … <n-1> … , 1, (n-1) } | $\frac{(n-2)}{2n}$ | 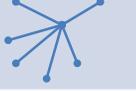 |
| 2. | (n-1)+(n-2) | {2, 2, … <n-2> … , 2, (n-1), (n-1)} | $\frac{(n-2)(n-3)}{n(2n-3)}$ | 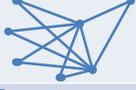 |
| 3. | (n-1)+(n-2) + (n-3) | {3, 3, … <n-3> …, 3, (n-1), (n-1), (n-1)} | $\frac{(n-3)(n-4)}{2n(n-2)}$ | 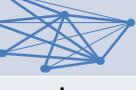 |
| | ⋮ | | | ⋮ |
| k. | (n-1)+(n-2) + (n-3) + … + (n-k) | {k, k, … <n-k> …, k, (n-1), … <k> …, (n-1)} | $\frac{(n-k)(n-k-1)}{n(2n-k-1)}$ | 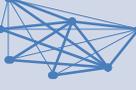 |
| | ⋮ | | | ⋮ |
| n. | $\binom{n}{2}$ | {(n-1), (n-1), … <n> …, (n-1)} | zero | 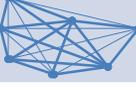 |

Figure 3. The table shows the highest sparsity index value (fourth column) and the corresponding degree sequence (third column) of a connected graph with given edge-count (second column) and $n$ number of nodes. The edge-count intervals are sequential and for a graph for which the edge-count falls in between two successive edge-counts (i.e., in two successive rows), the sparsity index falls within the corresponding range of highest sparsity values. It must be noted that as edge-count is an increasing sequence, the sparsity values decrease, until the value reaches zero for a complete graph.

### 2.3. Sparsity of edge-weights in weighted networks

In the previous section, section 2.2, we have dealt with the extreme values of sparsity index based on degree sequence, which apply equally on weighted and unweighted graphs. In this section, we specifically deal with the sparsity index based on edge weights of weighted graphs. First, we derive an expression of this index from its very definition based on Lorenz curve, given a distribution of the edge weights. Afterwards we look at some special cases in terms of bounding values and edge-weight distributions.

Let the edge-weight frequency distribution be denoted as in Table 1 in section 2.1. Also, let $T_1 = \sum_{i=1}^{k} w_i f_i$ denote the total weight of all edges of the graph. The Lorenz curve for the distribution of edge-weights of the graph can be drawn as in Figure 1.*ii*). The fractiles of weight frequencies (number of edges having a specific weight) are drawn along the horizontal axis and the cumulative proportion of edge-weights are drawn along the vertical axis. Thus, a point on the curve provides us the proportion of total edge-weights of the network (y-coordinate) that the specific fractiles (x-coordinate) of the edges possess.

To compute the sparsity geometrically from the Lorenz curve, we must compute the area under the Lorenz curve (area B as in Figure 1.*ii*)). We do so by piece-wise linear approximation of the Lorenz curve, following the standard technique in the literature [13]. We denote the area of a region by $\Delta$. An estimate of area B is thus obtained as follows:

$$\Delta(B) = \Delta(triangle\ \{(0,0), (\tfrac{f_1}{\binom{n}{2}},0), (\tfrac{f_1}{\binom{n}{2}}, \tfrac{w_1 f_1}{T_1})\}) + \Delta((k-1)\ trapeziums) = \tfrac{1}{2}\tfrac{f_1}{\binom{n}{2}}\tfrac{w_1 f_1}{T_1} + \Delta((k-1)\ trapeziums)$$

Area of the $i$-th trapezium is given by: $\frac{1}{2}\frac{f_i}{\binom{n}{2}}\left\{\frac{2\sum_{j=1}^{i-1}w_j f_j}{T_1} + \frac{w_i f_i}{T_1}\right\}, i = 2,3,…,k.$

Therefore, $\Delta(B) = \frac{1}{2}\frac{f_1}{\binom{n}{2}}\frac{w_1 f_1}{T_1} + \sum_{i=2}^{k}\frac{1}{2}\frac{f_i}{\binom{n}{2}}\left\{\frac{2\sum_{j=1}^{i-1}w_j f_j}{T_1} + \frac{w_i f_i}{T_1}\right\}.$

$$= \frac{w_1 f_1}{n(n-1)T_1}\{f_1 + 2f_2 + 2f_3 + \cdots + 2f_k\} + \frac{w_2 f_2}{n(n-1)T_1}\{f_2 + 2f_3 + \cdots + 2f_k\} + \cdots + \frac{w_i f_i}{n(n-1)T_1}\{f_i + 2f_{i+1} + 2f_{i+2} +$$
$$\cdots 2f_k\} + \cdots + \frac{w_{k-1} f_{k-1}}{n(n-1)T_1}\{f_{k-1} + 2f_k\} + \frac{w_k f_k}{n(n-1)T_1} f_k.$$

$$= \sum_{j=1}^{k-1} \frac{w_j f_j}{n(n-1)T_1}\{f_j + 2f_{j+1} + 2f_{j+2} + \cdots 2f_k\} + \frac{w_k f_k}{n(n-1)T_1} f_k.$$

$$= \sum_{j=1}^{k} \frac{w_j f_j}{n(n-1)T_1}\{f_j + 2f_{j+1} + 2f_{j+2} + \cdots 2f_k\}.$$

Hence, the sparsity index of a weighted network graph $G_w$ with weight frequency distribution as given in Table 1 is given by:

$$SI(G_w) = 1 - 2\Delta(B) = 1 - \frac{1}{\binom{n}{2}T_1} \sum_{j=1}^{k} w_j f_j \{f_j + 2f_{j+1} + 2f_{j+2} + \cdots + 2f_k\} \qquad (7)$$

It may be noted that in expression (7), $T_1$ truly represents the total edge-weight of all edges of the graph and hence $SI(G_w)$ also represents the Gini index of edge-weights of the graph. Having found the expression of sparsity index based on edge-weights of an undirected weighted graph, we proceed to investigate properties of this index for some specific graphs as follows.

    **I.**    A complete graph with equal edge weights

Let $w_i = w \;\forall\; i = 1, 2, \ldots, k$ with $w > 0$ be the unique edge-weight associated with all the $\binom{n}{2}$ edges of the graph. Note that $k = 1$ in this case. For, $w = 0$ indicates a graph with $n$ isolated vertices having total weight zero and $SI(G_w)$ is not defined as per our definition. Here, $T_1 = \sum_{i=1}^{k} w_i f_i = w\binom{n}{2}$ and $\Delta(B) = \frac{w\binom{n}{2}\binom{n}{2}}{2\binom{n}{2}T_1} = \frac{1}{2}$. Consequently, $SI(G_w) = 1 - 2\Delta(B) = 0$. The result is consistent with the fact that the sparsity of all equal quantities is zero.

    **II.**    A graph which has the first $(k-1)$ natural numbers as distinct edge-weights

Let $w_1 = 0$ with frequency $f_1$ and $w_i = i - 1$ with frequency $f_i$ where $f_i = 1 \;\forall\; i = 2, 3, \ldots, k$. This implies that the graph has $|E| = \sum_{j=2}^{k} f_j$ edges, each with distinct edge-weight and $\sum_{j=1}^{k} f_j = \binom{n}{2}$.

In this case, $T_1 = \sum_{i=1}^{k} w_i f_i = \sum_{i=2}^{k} w_i f_i = 1 + 2 + \cdots + (k-1) = \frac{k(k-1)}{2}$.

$$SI(G_w) = 1 - 2\Delta(B) = 1 - \sum_{j=1}^{k} \frac{w_j f_j}{\binom{n}{2}T_1}\{f_j + 2f_{j+1} + 2f_{j+2} + \cdots 2f_k\} = 1 - \sum_{j=1}^{k} \frac{w_j}{\binom{n}{2}T_1}\{1 + 2 + 2 + \cdots + 2\}.$$

$$= 1 - \sum_{j=1}^{k} \frac{w_j}{\binom{n}{2}T_1}\{1 + 2(k-j)\}.$$

$$= 1 - \frac{2}{\binom{n}{2}T_1} \sum_{j=1}^{k} \left(k - j + \frac{1}{2}\right) w_j \qquad (8)$$

As we compare expression (8) with expression (1), we observe that sparsity $SI(G_w)$ has been expressed in an equivalent form of sparsity index of the degree distribution $SI(G)$. Therefore, we have established equivalence in formulation of the two sparsity indices, namely, sparsity index of degree distribution of a simple network and a weighted undirected network with all distinct edge weights.

**Remark 1:** The expression (7) of sparsity index based on edge-weights of a network has been worked out with $T_1 = \sum_{i=1}^{k} w_i f_i$, and consequently the sparsity index coincides with the Gini index. However, this is not the only choice for $T_1$. For example, we may be interested to measure sparsity with respect to the total edge-weight obtained by considering the maximum edge-weight associated with any edge in the network, i.e., $T_1 = |E|w_n$. The $T_1$ here is obtained by considering edge weights as if every edge of the network has received the maximum edge weight possible, which is $w_n$, and that becomes the reference with respect to which the sparsity of the network is measured. The Lorenz curve differs in this case from what has been drawn in Figure 1.*ii)* and becomes more like the curve $OO'$ in Figure 1.*i)*.

**Remark 2:** Another important parameter in network parlance is *strength* of a node for weighted networks. *Strength* of a node is the total weight of all edges incident on that node, and is expressed as $s_i = \sum_j w_{ij}, i = 1,2,\ldots,n; j$ belongs to the set of neighboring nodes of node $i$. For non-negative integer edge-weights, a weighted graph can be treated as a multigraph as shown by Newman [29]. Therefore, the same expression (1) that holds for simple graphs can also hold for weighted graphs as weights are converted to degrees. Otherwise, an expression of the sparsity index can be obtained directly along the line of expression (1) based on strength sequence of the nodes of the network. However, the total strength $T_s$ of all nodes of the network now equals to twice the sum of all edge-weights of the network, as every edge-weight contributes equally to the two corresponding nodes the edge is attached to.

## 3. Edge-weight distribution and sparsity of networks

In this section, we are going to investigate sparsity of networks for different scenarios of edge-weight distribution. Based on empirical evidence, Barabási et al. [3] point out that large real-life networks self-organize into a scale-free state and display a power-law degree distribution. Also, the edge-weight distributions of a variety of large real-world networks have been shown to follow power-law distributions [8, 40]. Indeed, sparsity index is related with the exponent term of power law degree distribution of a network [18]. It should satisfy the same relationship with the exponent term of a power law edge-weight distribution, following the same logic. In this section, we attempt to uncover the behavior of sparsity index given certain distributions of edge-weights.

### 3.1. Network with binary weight values and its sparsity

Let us consider the simplest case of binary weight values of a network. Let $G_w$ denote a weighted network with edge-weight frequency distribution as in Table 2, in which a proportion $p, 0 \leq p < 1$, of the total possible edges assumes zero weight values and $1 - p$ proportion of the edges have unit weight values. We may note that a simple network can also be represented in this form, as its non-existing edges can assume zero weights, whereas all the existing edges take unit weights.

Table 2: Binary edge-weight distribution of a network

| Weight Values | Proportion of edges | Frequencies |
|---|---|---|
| $w_1 = 0$ | $p$ proportion of edges | $f_1 = \binom{n}{2} p$ |
| $w_2 = 1$ | (1- $p$) proportion of edges | $f_2 = \binom{n}{2}(1-p)$ |
| | | $T_1 = \sum_{i=1}^{2} w_i f_i = \binom{n}{2}(1-p)$ |

Therefore, for a network with binary weight values {0,1} distributed in proportion $p$ and $1 - p$, the sparsity index is given by:

$$SI(G_w) = 1 - \frac{1}{\binom{n}{2} T_1} \sum_{j=1}^{k} w_j f_j \{f_j + 2f_{j+1} + 2f_{j+2} + \cdots 2f_k\} = 1 - \frac{1}{\binom{n}{2}\binom{n}{2}(1-p)} \{\binom{n}{2}(1-p)\}^2 = p \qquad (9)$$

From the above formulation it is clear that $1 - p$ truly represents the edge density of the network. The idea of sparsity is complementary to density, which is aptly reflected in the above relationship, matching our intuition. A simple graph with, say 5% of the total possible edges as existing, has sparsity index 0.95 and is highly sparse.

The above result can be precisely stated in the form of Theorem 3.

**Theorem 3**: For a network with binary edge weights, if a proportion $(1 - p), 0 \leq p < 1$, of total possible edges are realized in the network, then its sparsity index is given by $p$.

### 3.2. Network with all distinct weight values and its sparsity

We consider a network $G_w$ in which a proportion $p, 0 \leq p < 1$, of the total possible edges assumes zero weights and the rest $(1 - p)$ proportion of the edges assume positive weight values distinct from each other. It indicates that $(1 -$

$p$) proportion of edges are realized in the graph. The corresponding edge-weight frequency distribution can be provided as in Table 3.

Table 3: edge-weight frequency distribution of a network with distinct weight values

| Weight Values | Proportion of edges | Frequencies |
|---|---|---|
| $w_1 = 0$ | $p$ proportion of the edges | $f_1 = \binom{n}{2} p$ |
| $w_2 \geq 1$ | $w_2, w_3, \ldots, w_k$ combined $1 - p$ proportion of the edges | $f_2 = 1$ |
| $w_3 > w_2$ | | $f_2 = 1$ |
| . | | |
| . | | |
| . | | |
| $w_j > w_{j-1}$ | | $f_j = 1 \; \forall \; j = 2, \ldots, k$ |
| $w_k > w_{k-1}$ | | $f_k = 1$ |
| | | $T_1 = \sum_{j=1}^{k} w_j f_j = \sum_{j=2}^{k} w_j$ |

Here, the total number of distinct weight values in the network is $k - 1 = \binom{n}{2}(1 - p)$ \hfill (10)

$$SI(G_w) = 1 - \frac{1}{\binom{n}{2}T_1} \sum_{j=1}^{k} w_j f_j \{f_j + 2f_{j+1} + 2f_{j+2} + \cdots 2f_k\} = 1 - \frac{2}{n(n-1)T_1} [\sum_{j=2}^{k-1} w_j \{1 + 2(k-j)\} + w_k f_k^2].$$

$$= 1 - \frac{2}{n(n-1)T_1} [\sum_{j=2}^{k-1} w_j \{2k + 1 - 2j\} + w_k] = 1 - \frac{2}{n(n-1)T_1} [(2k+1)(T_1 - w_k) - 2\sum_{j=2}^{k-1} j w_j + w_k].$$

$$= 1 - \frac{2}{n(n-1)T_1} [(2k+1)T_1 - 2k w_k - 2\sum_{j=2}^{k} j w_j + 2k w_k]$$

$$= 1 - \frac{2}{n(n-1)T_1} [(2k+1)T_1 - 2\sum_{j=2}^{k} j w_j] \hfill (11)$$

Now, $\sum_{j=2}^{k} j w_j \leq k \sum_{j=2}^{k} w_j = k T_1$ \hfill (12)

Combining expressions (11) and (12), we get

$$SI(G_w) \leq 1 - \frac{2}{n(n-1)T_1} [(2k+1)T_1 - 2k T_1] = 1 - \frac{2}{n(n-1)} \hfill (13)$$

The expression (13) holds for $n > 2$ and provides us an upper bound of the sparsity index. This bound depends only on the number of nodes of the network, and, is independent of the density $1 - p$ of the network. We also find a lower bound of the index as follows.

$SI(G_w) \geq p \Leftrightarrow 1 - \frac{2}{n(n-1)T_1}[(2k+1)T_1 - 2\sum_{j=2}^{k} j w_j] \geq p$, follows from equation (11)

$\Leftrightarrow \frac{4 \sum_{j=2}^{k} j w_j}{n(n-1)T_1} \geq \frac{2(2k+1)}{n(n-1)} - (1-p) \Leftrightarrow \frac{\sum_{j=2}^{k} j w_j}{T_1} \geq \frac{2k+1}{2} - \frac{n(n-1)(1-p)}{4} = k + \frac{1}{2} - \frac{k-1}{2}$, substituting $\binom{n}{2}(1-p) = k - 1$

$\Leftrightarrow \frac{\sum_{j=2}^{k} j w_j}{T_1} \geq 1 + \frac{k}{2} \Leftrightarrow \frac{\sum_{j=2}^{k} j w_j}{\sum_{j=2}^{k} w_j} \geq 1 + \frac{k}{2}.$ \hfill (14)

But, $1 \leq w_2 < w_3 < \cdots < w_{k-1} < w_k$. This implies that $\frac{\sum_{j=2}^{k} j w_j}{\sum_{j=2}^{k} w_j} > \frac{\sum_{j=2}^{k} j}{k-1} = \frac{1}{k-1} \left[ \frac{k(k+1)}{2} - 1 \right] = 1 + \frac{k}{2}.$ \hfill (15)

Combining relationships (14) and (15) we get a strict lower bound of $SI(G_w)$, i.e., $SI(G_w) > p$.

The above result is summarized in the form of Theorem 4.

**Theorem 4**: For a network $G_w$ with number of nodes $n > 2$, let the number of actual edges be $1 - p$ proportion of the total possible edges, and let the edge-weights be all distinct. The sparsity index $SI(G_w)$ based on edge-weights of the network satisfies: $p < SI(G_w) \leq 1 - \frac{2}{n(n-1)}$.

**Remark:** Even for a complete graph for which $p = 0$, $SI(G_w)$ can be close to 1, indicating that edge density does not play a part in counting sparsity of the edge-weights.

### 3.3. A complete network with two distinct equally distributed edge-weights and its sparsity

Finally, we consider a special type of a complete network, namely, a complete network with two distinct edge-weight values distributed equally among its edges, i.e., half of the edges of the network assume a particular weight value and the rest half receive another weight value distinct from the other one. The weight distribution is described in Table 4.

Table 4: Edge-weight distribution of a complete network with only two weight values

| Weights | Proportion of edges | Frequencies |
|---|---|---|
| $w_1 \geq 1$ | ½ of the edges | $f_1 = \binom{n}{2}/2$ |
| $w_2 > w_1$ | ½ of the edges | $f_2 = \binom{n}{2}/2$ |

$$T_1 = \sum_{i=1}^{k} w_i f_i = \frac{\binom{n}{2}}{2}(w_1 + w_2)$$

In this case,

$$SI(G_w) = 1 - \frac{1}{\binom{n}{2}T_1}\sum_{j=1}^{k} w_j f_j \{f_j + 2f_{j+1} + 2f_{j+2} + \cdots 2f_k\} = 1 - \frac{1}{\binom{n}{2}T_1}\{w_1 f_1^2 + w_2 f_2^2 + 2w_1 f_1 f_2\}$$

$$= 1 - \frac{1}{\binom{n}{2}\binom{n}{2}\frac{1}{2}(w_1+w_2)}\left[w_1 \frac{\binom{n}{2}}{2}\frac{\binom{n}{2}}{2} + w_2 \frac{\binom{n}{2}}{2}\frac{\binom{n}{2}}{2} + 2w_1 \frac{\binom{n}{2}}{2}\frac{\binom{n}{2}}{2}\right] = 1 - \frac{3w_1+w_2}{2(w_1+w_2)} = \frac{w_2-w_1}{2(w_1+w_2)}.$$

Let $w_2 = \alpha w_1$; $\alpha > 1$.

Then $SI(G_w) = \frac{w_1(\alpha-1)}{2(\alpha+1)w_1} = \frac{1}{2} - \frac{1}{\alpha+1} < \frac{1}{2}$.

For large $\alpha$, $SI(G_w) \rightarrow \frac{1}{2}$.

### 4. Community Detection and Sparsity of Weighted Networks

In this section, we are interested in exploring the use of weighted sparsity index towards network analysis. A major area within network analysis is devoted to reveal functions and organizations of networks through the discovery of community structures. Although there is a plethora of community detection algorithms which can be applied to weighted networks, each one of them works with full efficiency within its specific area of application, i.e., domain context, underlying assumptions and the specific objective to be achieved. Moreover, there are several approaches to detect communities within networks. For example, Palla et al. [32] use an adjacent $k$-clique based approach, in which a $k$-clique community is defined as a union of all $k$-cliques which can be reached from each other through a series of adjacent $k$-cliques (two $k$-cliques are adjacent if they share $(k - 1)$-many nodes). Another approach is to identify natural groupings of related nodes in the network to optimize a modularity function [30]. A fast label-propagation based approach [37] allows the labels of nodes to propagate at every step of the process such that each node adopts the label most of its neighbors have at that point. The process stops as the nodes do not change labels. This approach is different from the earlier ones, as it only uses knowledge about local network structure or local modularity. Several algorithms [2, 14, 36] are based on *star subgraphs*, which generate node-cover of the network by dense subgraphs that are star shaped [2]. As network community detection is an active area of research, it may be worthwhile to explore applicability of the proposed index towards detecting communities in weighted sparse networks.

An overlapping community detection algorithm for simple graphs has been proposed by Goswami et al. [18]. It utilizes the property of sparsity index to detect disparity in degrees of nodes to produce groupings of nodes which are more homogeneous among themselves in terms of connectivity. We examine suitability of extending this sparsity index based greedy, node-oriented overlapping community detection method to weighted networks. We may recollect that this method emphasizes on allowing the communities to evolve naturally, without any prior specification of community size or number of communities and generates overlapping communities. Moreover, it has got three main phases, namely, **splitting, reconstruction and final tuning**. In a weighted network, the effect of edge-weights can be transferred to the nodes by computing node-strength. Strength of a particular node of a network is the total weight of all edges incident on that node. Just as a simple network has its ordered degree sequence, a weighted network can have its ordered strength sequence. Therefore, the first modification that can be introduced in the method is in its **splitting** stage, in which star subgraphs around nodes of highest strength, instead of highest degree, are repeatedly extracted from the network, until the network degenerates into components all of which are regular subgraph (or isolated nodes). The sparsity index used in this phase is to be based on node strength and the quantity with respect to which it is measured, i.e., $T_1$ is taken as $ns_{max}$, where $s_{max}$ denotes the maximum strength of a node in the graph. Note that in computing sparsity index for the degree distribution of unweighted graph, $T_1$ has been taken as $n(n-1)$, the potential total degree of the graph. In the **reconstruction** phase of the method, candidate communities are formed. At this point, one may utilize weighted sparsity index $SI(G_w)$ as a criterion to choose the candidate community from possible combinations of regular graph components and extracted star subgraphs. As $SI(G_w)$ represents the variability in the distribution of edge-weights of the weighted graph $G_w$, the method attaches preference to the subgraph, which has a more homogeneous edge-weight distribution, to be a candidate community. In reality, a "good" community within an interaction network is that part of the network in which entities are more connected and interact more consistently and uniformly with each other. The third phase is about **tuning** the set of candidate communities to get a more meaningful set of final communities. It does so by merging small communities, wherever possible, to the larger ones in case there is a significant overlap, removing redundancy and so on. We introduce a threshold at this phase, as a user supplied parameter, to denote the percentage of overlap beyond which two candidate communities are to be merged. This phase largely follows its predecessor otherwise. The new method thus formed has been described in section 4.1.

In dealing with weighted networks for community detection, we have noted in the literature that the edge weights have been accounted for in several ways. One method [24] argues that if the edge weight between two nodes is high enough, they should belong to the same community. The method is driven by this principle as it starts with identifying the nodes associated with the highest edge-weights in the network and progressively builds communities by optimizing "conductance" function. Another method [7] builds the initial community as the star subgraph around the node of highest strength and expands (or contracts) the community based on criteria defined primarily on "belonging degree" (belonging degree of a node to a community is the fraction of its node-strength due to its neighbors in the community). The DClustR method proposed by Perez-Suarez et al. [34] uses a "relevance" score defined on every node of the graph, to decide whether to include the weighted star sub-graph ("ws-graph" as the authors name it) around that node as an initial cluster. The initial set of clusters is improved upon by trying to reduce the number of clusters and the overlapping among them using suitable strategies to obtain the final result.

### 4.1. Description of the proposed method

It must be mentioned that the proposed method does not require any user supplied input in terms of the number or size of the communities. The only parameter that the method makes use of at the final tuning stage is a threshold to determine to what extent the method is to allow overlap among the final list of communities that are generated as output. The steps of the proposed method are outlined as follows:

**(i) Splitting**

The aim is to split the graph in such a way that, on one hand, induced subgraphs around nodes of highest strength start getting gathered into one list, say $L$, and on the other, the remaining components if found regular (including isolated nodes) are collected in another list, say $C$. The whole graph is split into components in lists $L$ and $C$.

- Given the ordered strength sequence $\{s_1, s_2, ..., s_n\}$, the node, say $v$, with the highest strength is chosen. In case of a tie in such selection, the node with a larger clustering coefficient is used. If the tie persists, an arbitrary $v$ is chosen from the nodes with the highest strength and equal clustering coefficients. Clustering coefficient of a node $v$ is defined by Watts and Strogatz [43] as the proportion of triangles connected to node $v$ compared to the number of triplets (open or closed triangles) associated with node $v$.
- The graph $G_w$ is split around node $v$ to produce two parts: a weighted star sub-graph centered around node $v$, say $G^*_{w_v}$ and $G_w \setminus G^*_{w_v}$.
- Two initially empty lists $L$ and $C$ are maintained. The weighted induced subgraph comprising of nodes of the star subgraph $G^*_{w_v}$, i.e., $\{v\} \cup N(v)$ is inserted into list $L$. Here $N(v)$ denotes the neighbors of node $v$.
- The graph $G_w \setminus G^*_{w_v}$ may remain connected or may be broken down into multiple connected components. Each connected component is checked for regularity, i.e., whether its Gini index value is zero. If so, the component is inserted into the list $C$.
- The process is repeated until the graph $G_w$ is decomposed into components each with a zero Gini index, i.e., each component of list $C$ is a regular graph or an isolated node.

**(ii) Reconstruction**

The goal of reconstruction is to construct candidate communities out of fused subgraphs from lists $L$ and $C$, using edge-weight sparsity index as the criterion for selection of the candidate community.

- Every member subgraph of the list $L$ is matched, by taking union, against all the component subgraphs of list $C$ to check if the resulting union is a connected subgraph. If no connected subgraph can be formed thus, the member subgraph stays in the list $L$ and the next member of list $L$ is processed. If multiple connected subgraphs are formed with different components in list $C$, sparsity index based on edge weights is computed for each of them and the one with the least sparsity index replaces the corresponding component in list $C$. Any contention in such selection is resolved by giving preference to the component having fewer edges. The member subgraphs of list $L$ are processed on a First in First out (FIFO) basis.
- The process is repeated for every member of list $L$.
- The component subgraphs of list $C$ together with member subgraphs staying back in list $L$, if any, comprise the candidate list of communities.

**(iii) Final Tuning**

The list of candidate communities is sieved through this tuning process to obtain the final list of communities by eliminating trivial groupings and by merging communities which have a significant overlap.

- Any candidate community, which happens to be a proper subset of any other candidate community, is eliminated from the list.
- If a candidate community of size two or three, is found to share its nodes with any other larger candidate community, it is merged with that one through union operation (in case of multiple such options, an arbitrary one is chosen).
- Any *redundant* candidate community is removed from the list. By redundancy, we mean that the union of the rest of the candidate communities, excluding the present one, can cover the entire graph).
- If any two candidate communities have a proportion, say $h$, of their nodes in common, they are merged. $h$ can be a predetermined threshold depending on the current context of the community detection task.
- The final list of communities is generated at the end of processing the above steps.

It can be noted that the algorithm produces the same set of communities of a graph in different executions, except for a rare possibility of resolving ties through arbitrary selection. In the splitting stage, the order of selection is of no consequence for member subgraphs of list $L$. However, the ordering of member subgraphs in list $L$ may impact the outcome of the reconstruction process. Also, the sparsity index of the strength sequence of the graph is checked against a preset threshold, to ensure that the graph under consideration is sparse enough for the community detection exercise to be meaningful.

The **computational complexity** of the present method is $O(n^2)$ for sparse graphs (number of edges is linear in number of nodes of the graph [10]), $n$ being the number of nodes of the graph. The detailed computation has been shown by the predecessor of the present algorithm, i.e., algorithm for simple networks [18]. Although the sparsity index of edge-weights has been computed additionally for the present algorithm, its computational cost does not exceed the mentioned one.

### 4.2. Result

The present method of overlapping community detection is a node-based method. For evaluation, the method has been applied on synthetic and real-life networks as described in subsequent sub-sections. For measuring quality of the detected communities, we have used the most commonly used internal quality metric, a modularity-based measure extended for overlapping communities [39]. The quality metric is defined as follows:

*Extended Modularity of overlapping community structure* $Q_o = \frac{1}{S}\sum_{c=1}^{k}\sum_{i,j=1}^{n}\alpha_{ci}\alpha_{cj}(a_{ij} - \frac{s_i s_j}{S})$, (15) where $k$ denotes the number of communities; $\alpha_{ci} = \frac{s_{ci}}{\sum_{c=1}^{k}s_{ci}}$, with $s_{ci} = \sum_{j \in c} a_{ij}$ denotes the strength of the $i$-th node in the community $c$. Also, $0 \leq \alpha_{ci} \leq 1\ \forall\ c = 1,2,\ldots,k; i = 1,2,\ldots,n$. Moreover, $\alpha_{ci} = 1$ if the node $i$ belongs to the community $c$ alone and $\alpha_{ci} = 0$ if node $i \notin$ community $c$. The quantity $S = \sum_{i=1}^{n}\sum_{j=1;j\neq i}^{n} a_{ij}$ denotes the total edge-weight of the network.

#### 4.2.1. Synthetic Weighted networks

The first network on which we apply the weighted sparsity-index based community detection method is a synthetic edge-weighted network used by Perez-Suarez et al. [34] to illustrate their own method of overlapping community detection in weighted graphs. Notably, the same network has been used to illustrate the original, unweighted version of the present method. Although the two methods just mentioned have significant differences in criteria of community formation, the detected community structures match almost entirely. It is worth mentioning that the former method did consider the edge-weights of the network, while the later ignored them completely. As we apply the newly constructed algorithm, along the line described in the section 4.1, on the same network, it results in the same set of communities. As the new splitting process is driven by node strength rather than node degrees, the order in which the star subgraphs are extracted from the network, vary. The order of selection of the highest degree nodes is ("18","8","4","12"), whereas that of the highest strength nodes is ("8","4","18","12"). The present reconstruction process considers the edge-weights using weighted sparsity index. Nevertheless, the result remains the same. The identified communities have been marked in Figure 4. The result could have been different had the edge weights been distributed in a different manner in this graph. The reason for having the same set of communities as generated by the unweighted version of the algorithm can be attributed primarily to the distribution of edge-weights in this particular network.

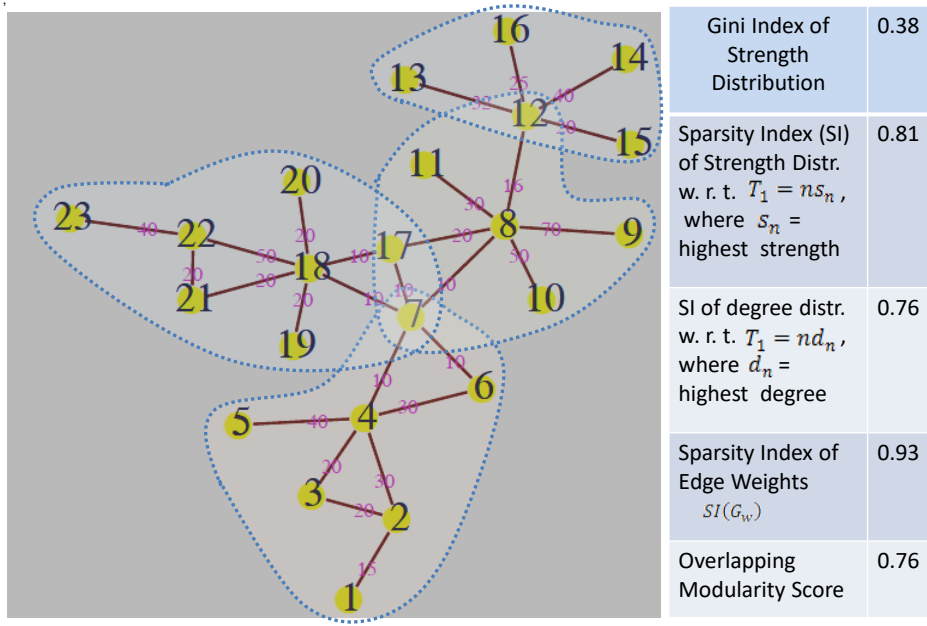

**Figure 4.** Communities as detected by the present algorithm on the example graph [34], along with relevant sparsity indices and quality measures.

### 4.2.2. Real Life Networks

The sparsity index-based community detection method for simple networks has been applied by its authors on two well-known benchmark networks, namely, Karate club network and Dolphin social network. Both of these networks are unweighted. Therefore, to apply the modified algorithm for weighted networks, we choose to replace them with real life networks, which are weighted and sparse.

I. **The collaboration dataset on Network Science** [29] depicts a network of 1589 scientists working on topics in network science and related disciplines, as compiled by M. E. J. Newman in 2006. The largest connected component, consisting of 379 scientists and 914 pair-wise collaborations, is going to be used for our experiment. This network science graph has Gini index and sparsity index in terms of node-strength as 0.47 and 0.96, respectively and it has the weighted sparsity index $SI(G_w) = 0.993$.

The network science collaboration network generates 51 communities initially as the proposed algorithm is applied. The resulting modularity score for overlapping communities is computed to be 0.64. As communities with significant overlaps (more than 26% common nodes) are merged, we get 32 communities with a modularity score of 0.82. The merging operation takes away large overlaps among communities and fuses some of the communities, insignificant in size, with their larger counterparts. The community structure thus obtained is portrayed in Figure 5.

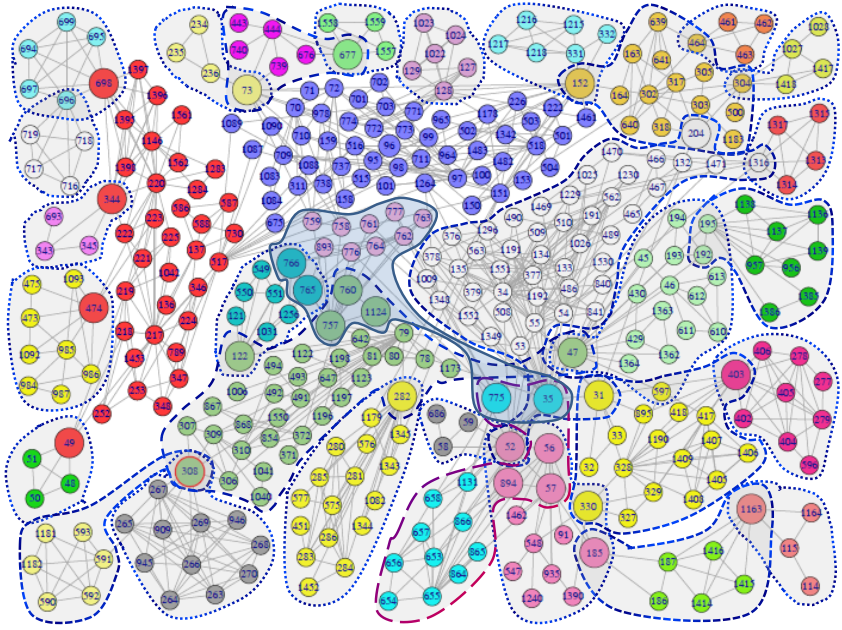

**Figure 5.** The community structure of Network Science collaboration network, as identified by the present algorithm. The enlarged nodes indicate shared nodes.

In the resulting grouping of nodes, a node is shared by at most three communities and there are six such nodes, namely, nodes 35, 52, 73, 152, 282 and 308. Among these six cross-disciplinary researchers are H Jeong (35), A Vespignani (152), J Kleinberg (308).

In this context it may be interesting to note a couple of results obtained by other researchers on the same network, the network science collaboration network.

i. Refer to the work of Chen et al. [7] on proposing an overlapping community detection algorithm of weighted networks. Table 2 of this article mentions that the modularity score for overlapping communities as obtained by CONGA, an overlapping version of the well-known label propagation algorithm, is 0.75; that of CFinder, a $k$-clique percolation based algorithm, is 0.42 and that of the authors' own algorithm is 0.85. In comparison with these benchmark community detection algorithms, the proposed algorithm in the present article is placed competitively, provided modularity is the quality criterion.

ii. The work of Held and Kruse (Figure 6 of [21]), shows the relationship between the number of clusters and modularity of different partitions of the network science dataset by four different algorithms. Majority of these algorithms get the maximum value of the modularity around 20 to 30 clusters. As the present algorithm completes its final tuning process, 32 overlapping communities emerge. Notably, the community structure reported by Chen et al. [7] has 31 overlapping communities.

**II.** This benchmark dataset has been taken from the **les miserable** dataset as compiled by Knuth [22], a co-appearance network of characters in the novel les Miserables by V. Hugo. A pair of nodes in the network is linked if the corresponding characters appear together at least once in one or more chapters of the book. There are 77 nodes and 254 edges in the network and the edge-weights range between 1 and 31. The Gini index and sparsity index of node-strength of this network are respectively 0.6 and 0.95 and the sparsity index of edge-weights is also 0.95.

As the present algorithm is applied on this network, nine overlapping communities are formed initially. At the final tuning stage, communities with 50% or more overlapping nodes are merged and this results in five communities as shown in Figure 6. The largest community is centered on node 12, Jean Valjean, the protagonist of the novel. It also includes his persecutor, the police officer Javert, node no 28. The next largest community includes Marius, node no

56, the revolutionaries, node nos. 58 to 68 and two young children, node nos. 74 and 75, among a few others. The third community consists of the community of students, node nos. 17 to 23 and includes Fantine, node no 24 and Cosette, node no. 27. Interestingly, in this overlapping community structure based on node-strength, Marius appears in four out of five communities and Thenardier, node no. 26, appears in three. The community centered on Bishop Myriel, node no. 1, shares an overlap of four characters with the largest community centered on Valjean. The overlapping modularity score of the recognized grouping is 0.49.

As to the quality of the community structure, we refer to the works of Medus et al. [26] and Zalik [46], in which this network has been used for validating the community detection algorithms proposed by the respective authors. Medus et al. point out that the best community structure has a modularity of 0.546 and corresponds to five communities (centered on Jean Valjean, Javert, Marius, Fantine and bishop Myriel). However, the algorithm proposed by Zalik recognizes six communities and the modularity of the obtained partition is 0.501 (with parameter $k = 1$). It is to be noted that both referred algorithms are distinct partition-based, i.e., non-overlapping algorithms. The grouping generated by the CFinder algorithm has a modularity score of 0.436 [46]. The modularity of the proposed algorithm is not the best, but it is placed competitively with those of other similar algorithms.

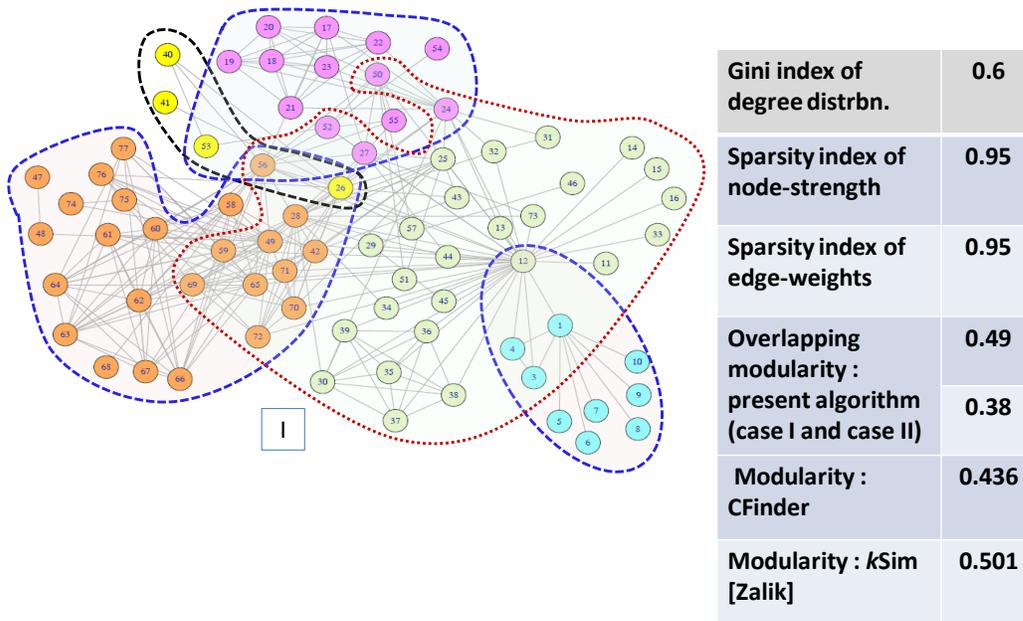

| | |
|---|---|
| Gini index of degree distrbn. | 0.6 |
| Sparsity index of node-strength | 0.95 |
| Sparsity index of edge-weights | 0.95 |
| Overlapping modularity : present algorithm (case I and case II) | 0.49 |
| | 0.38 |
| Modularity : CFinder | 0.436 |
| Modularity : $k$Sim [Zalik] | 0.501 |

**Figure 6. Community discovery in the les Miserables network by the present algorithm based on node-strength. Communities are color-coded and marked on the diagram.**

It is important to highlight that the present method approaches community discovery based on strength of nodes, rather than their degrees. Hence, the central characters of the recognized communities may differ from those obtained by the other algorithms with which our results have been compared. We recognize that edge-weights in this network denote the number of times the corresponding characters co-appear; detecting the inherent groupings of the network based on node-strength would mean that the frequency of co-appearance also receives due concern in the detection process.

We stop short of testing the present method further using more datasets and comparisons with existing algorithms, to maintain the main focus of this paper. Our primary intention has been to illustrate an application of the sparsity indices, described in this paper, in devising a community detection method for weighted sparse networks. The application on a synthetic network and the benchmark real life networks show that the method works at par with some of the well-known node-based community detection methods, in terms of accuracy and computational complexity.

## 5. Conclusion

For a network graph, sparsity is a property to indicate the inherent variability of a certain quantity associated with the graph. Among different measures of sparsity, in this article we have focused on a specific measure called sparsity index, defined along the line of Gini index, from Lorenz curve. Sparsity index is a summary measure and is normalized. It was defined originally on the ordered degree sequence of a network graph. A theoretical result worked out in the present article shows that for any graph which has a tree structure with no less than three nodes, its sparsity index lies between the sparsity index of a Hamiltonian path and that of a star graph with the same number of nodes. The gradual evolution of tree structures from Hamiltonian path to star graph, in ascending order of their sparsity indices, is demonstrated. The result is extended to cover the entire range of connected networks having a minimum of three nodes. Instead of having a fixed number of edges, sequential intervals of edge-counts are taken, to cover all possible connected graphs (having tree structures to complete graphs); the corresponding degree sequences and the sparsity index range are computed. There is scope to adopt similar exercise in future to find out further properties of this index in connection with other network parameters, especially, to link it to algorithms which deal with sparse networks.

Throughout this article, sparsity index is used in connection with different quantities concerning a weighted network, like its degree sequence, strength sequence, edge-weights etc. Our intention here is to emphasize that each of these measures reveal different aspects of the network and the knowledge thus obtained can be very useful for network analysis. And yet the measures are relatively inexpensive to compute (to the order of $nlogn$). Particularly, the community detection algorithm that we have proposed in this context not only makes use of each these sparsity indices (it uses sparsity index of strength sequence, degree sequence and edge-weights), but also it is a promising overlapping community detection algorithm. It turns out that the computational complexity of the proposed method is at par with other well-known methods of overlapping community detection for networks. The accuracy of the results, tested on a few datasets, is also found to be at par with some of the existing methods. In order to establish the method fully, more experimental results are needed to be incorporated. But we have not elaborated the results further to stay in tune with the main focus of the paper, which is to extend the sparsity index beyond degree sequence and provide some results, both theoretical and experimental, for showing its utility. We intend to work more on the community detection method in future to enhance its efficiency to make it a full-blown overlapping community detection method for networks.

Alongside the structure of a network the content has been brought into focus by extending the sparsity measure to edge-weights. Given the frequency distribution of edge-weights of a network, an expression is derived for the sparsity index, which immediately reflects the inherent heterogeneity of edge weights of the network. The relationship between heterogeneity of edge weights and network topology has been explored [31] and it has been shown that for mobile data networks the two are strongly correlated. The authors show that the strongest edges lie inside dense structures of the networks, while the weak ones act as connectors between densely organized groups. Just as degree distribution in a wide variety of real life networks display power law characteristics, so does edge weight distribution in certain cases [8, 40]. The distribution of edge weights for complex networks is a less charted area and there is much to explore, both from a theoretical and an application-oriented perspective, regarding how such a distribution relates with other characteristics of the network. The sparsity index on edge weights may be of crucial help in this direction.

## 6. Acknowledgements

This work has been financially supported by the Department of Science and Technology, Government of India through a research grant under the Women Scientist Scheme – A, vide reference number ET28/2017, as part of the project titled "characterization and analysis of interaction networks". We are deeply indebted to late Prof. C. A. Murthy for his contribution towards producing this paper. With a sense of gratitude, we fondly remember his insight into making instructive suggestions and critical comments to enhance quality of this article.